\documentclass[twocolumn,superscriptaddress,showpacs,pra]{revtex4}
\usepackage{epsfig}
\usepackage{color}
\usepackage{delarray}
\usepackage{amsmath, amssymb}
\usepackage{bm}

\begin{document}
\title{Shapes and Statistics of the Rogue Waves Generated by Chaotic Ocean Current\footnote{This article can be cited as: Bayindir, C., 2016, ``Shapes and Statistics of the Rogue Waves Generated by Chaotic Ocean Current", ISOPE-2016, Rhodes, Greece.}}

\author{Cihan Bay\i nd\i r}
\email{cihan.bayindir@isikun.edu.tr}
\affiliation{Department of Civil Engineering, I\c s\i k University,  \.{I}stanbul, Turkey}

\begin{abstract}

In this study we discuss the shapes and statistics of the rogue (freak) waves emerging due to wave-current interactions. With this purpose, we use a simple governing equation which is a nonlinear Schr\"{o}dinger equation (NLSE) extended by R. Smith (1976). This extended NLSE accounts for the effects of current gradient on the nonlinear dynamics of the ocean surface near blocking point. Using a split-step scheme we show that the extended NLSE of Smith is unstable against random chaotic perturbation in the current profile. Therefore the monochromatic wave field with unit amplitude turns into a chaotic sea state with many peaks. By comparing the numerical and analytical results, we show that rogue waves due to perturbations in the current profile are in the form of rational rogue wave solutions of the NLSE. We also discuss the effects of magnitude of the chaotic current profile perturbations on the statistics of the rogue wave generation at the ocean surface. The extension term in Smith's extended NLSE causes phase shifts and it does not change the total energy level of the wave field. Using the methodology adopted in this work, the dynamics of rogue wave occurrence on the ocean surface due to blocking effect of currents can be studied. This enhances the safety of the offshore operations and ocean travel.

\end{abstract}
\maketitle

\section{Introduction}
Rogue (freak) waves are the waves with a height more than 2-2.2 times the significant wave height in the ocean wave field. They present a danger to the safety of marine operations and life and their result can be catastrophic and costly (Kharif and Pelinovsky, 2003; Bay\i nd\i r, 2015). The understanding of chaotic rogue wave dynamics is a must for the safety of the marine travel and the offshore operations in stormy conditions. There are few known mechanisms in ocean which cause rogue waves formation. These are focusing of the nonlinear wave interactions such as collusion of breathers, frequency modulated wave groups due to geometrical and dispersion effects, atmospheric forcing and wave-current interactions in strong currents.   

\
In this paper we consider the generation of the rogue waves in the strong wave-current interactions regions such as the Agulhas current of the southwest Indian ocean or the Gulf Stream current in the Atlantic ocean (Kharif and Pelinovsky, 2003). With this motivation, we use a simple governing equation first proposed by Smith (1976) which is a nonlinear Schr\"{o}dinger equation (NLSE). The extended NLSE of Smith accounts for the effects of current gradient on the dynamics of the ocean surface. Implementing a split-step numerical scheme we show that the extended NLSE of Smith is unstable against random chaotic perturbation in the current profile. Therefore the initial monochromatic wave field turns into a chaotic sea state with many apparent peaks at later times. We show that those rogue waves can be described by the rational rogue wave solutions of the NLSE. Additionally we discuss the effects of magnitude of the chaotic current profile perturbations on the statistics of the rogue wave generation. We show that extension term in Smith's extended NLSE introduces phase shifts to the solutions and it does not alter the total energy of the wave field. Using the methodology discussed in this paper, the dynamics of rogue wave on the ocean surface due to blocking effect of currents can be analyzed which can be used for enhancing the safety of the maritime operations and ocean travel.

\section{Rogue Waves in a Chaotic Wave-Current Field}

\noindent NLSE is widely used in many branches of the applied sciences and engineering. It describes various phenomena including but not limited to pulse propagation in optical fibers, Bose-Einstein condensation, quantum state of a physical system and quantum control just to name a few. It is also possible to model the dynamics of weakly nonlinear deep water ocean waves by the NLSE (Zakharov, 1968). In order to study various dispersion, focusing, nonlinearity and other effects, many modified and extended versions of the NLSE are proposed. In order to examine the effects of the ocean current gradient on the nonlinear wave field dynamics in the vicinity of the blocking point, Smith derived the extended NLSE equation (Smith, 1976; Kharif and Pelinovsky, 2003) in the form
\begin{equation}
i\psi_t + \alpha \psi_{xx} +  \beta \left|\psi \right|^2 \psi - \gamma x\left|\frac{dU}{dx} \right|\psi =0
\label{eq01}
\end{equation}
where $x, t$ is the spatial and temporal variables, $i$ denotes the imaginary number, $\alpha, \beta, \gamma $ are constants, $\psi$ is complex amplitude, $U$ is the current speed and $dU/dx$ is the current gradient calculated in the vicinity of the blocking point (Kharif and Pelinovsky, 2003). The constants in front of the terms of the extended NLSE of Smith in the original form are different, but they can be reduced to the form given in Eq.\,(\ref{eq01}) by simple scalings and Galilean shifts (Smith, 1976; Kharif and Pelinovsky, 2003). The extended NLSE of Smith in Eq.\,(\ref{eq01}) is also used for investigating the nonlinear effects near caustics (Peregrine and Smith, 1979). A transformation of the form 
\begin{equation}
\begin{split}
\psi(x,t)= \widetilde{\psi} & \left (x+ \alpha \gamma t^2 \left|\frac{dU}{dx} \right|, t \right) \\
& .\exp{\left( -i \left[\gamma xt \left|\frac{dU}{dx} \right|+ \alpha \gamma^2 \frac{t^3}{3} \left|\frac{dU}{dx} \right|^2 \right] \right)}
\label{eq02}
\end{split}
\end{equation}
transforms the extended NLSE of Smith to the cubic NLSE of the form
\begin{equation}
i\widetilde{\psi}_t + \alpha \widetilde{\psi}_{xx} + \beta  \left|\widetilde{\psi} \right|^2 \widetilde{\psi} =0
\label{eq03}
\end{equation}
for which the rational rogue wave solutions becomes obvious (Akhmediev et.al., 2009a; Akhmediev et.al., 2009b; Akhmediev et.al., 2011). For $\alpha=0.5, \beta=1 $ the cubic NLSE can be written as 
\begin{equation}
i\widetilde{\psi}_t + 0.5 \widetilde{\psi}_{xx} + \left|\widetilde{\psi} \right|^2 \widetilde{\psi} =0
\label{eq035}
\end{equation}
 One of the most early forms of the rational soliton solution of the NLSE is the Peregrine soliton which can describe the first order rogue wave. It is given by
\begin{equation}
\widetilde{\psi}_1=\left[1-4\frac{1+2it}{1+4x^2+4t^2}  \right] \exp{[it]}
\label{eq04}
\end{equation}
where $x$ is the space and $t$ is the time parameter. It is shown that Peregrine soliton is only a first order rational soliton solution of the NLSE (Akhmediev et.al., 2009b). The second order rational soliton solution is given by Akhmediev et.al. (2009b) as
\begin{equation}
\widetilde{\psi}_2=\left[1+\frac{G_2+it H_2}{D_2}  \right] \exp{[it]}
\label{eq05}
\end{equation}
where
\begin{equation}
G_2=\frac{3}{8}-3x^2-2x^4-9t^2-10t^4-12x^2t^2
\label{eq06}
\end{equation}
\begin{equation}
H_2=\frac{15}{4}+6x^2-4x^4-2t^2-4t^4-8x^2t^2
\label{eq07}
\end{equation}
and
\begin{equation}
\begin{split}
D_2=\frac{1}{8} [ \frac{3}{4} & +9x^2+4x^4+\frac{16}{3}x^6+33t^2+36t^4+\frac{16}{3}t^6 \\
&-24x^2t^2+16x^4t^2+16x^2t^4 ]
\label{eq08}
\end{split}
\end{equation}

Third and higher order rational solutions of the cubic NLSE and a hierarchy of obtaining those rational soliton solutions that relies on the Darboux transformations are given by Akhmediev et.al. (2009b). Many simulations and experiments have revealed that hydrodynamic rogue waves are in the forms of these first (Peregrine) and higher order rational solutions of the NLSE (Akhmediev et.al., 2009a; Akhmediev et.al., 2009b; Akhmediev et.al., 2011). The cubic NLSE given in Eq.\,(\ref{eq035}) is invariant under the scale, phase and Galilean transformations (Bay\i nd\i r, 2015)
\begin{equation}
\widetilde{\psi}(x,t) \rightarrow B \widetilde{\psi}(Bx,B^2t), \ \ \ B \in \Re^+ 
\label{eq09}
\end{equation}

\begin{equation}
\widetilde{\psi}(x,t)  \rightarrow  \exp{[ic]} \widetilde{\psi}(x,t), \ \ \ c \in \Re 
\label{eq10}
\end{equation}

\begin{equation}
\widetilde{\psi}(x,t) \rightarrow  \widetilde{\psi}(x-Vt,t) \exp{[iVx-iV^2t/2]}, \ \ \ V \in \Re
\label{eq11}
\end{equation}
Using Eq.\,(\ref{eq02}), Eq.\,(\ref{eq03}) and Eq.\,(\ref{eq11}) one can realize that a transformation of the form
\begin{equation}
 \left| \widetilde{\psi} (x,t) \right| \rightarrow  \left| {\psi}\left(x-\left[V-\frac{t}{2} \left|\frac{dU}{dx} \right| \right]t,t \right)  \right|, \ \ \ V \in \Re
\label{eq12}
\end{equation}
leads to a progressive solution of the extended NLSE of Smith. This shows that solutions of the NLSE can be transmitted or reflected by the current depending on the sign of the celerity $\left[V-\frac{t}{2} \left|\frac{dU}{dx} \right| \right]$.

A chaotic sea state involves a full range of the wave spectrum with many spectral components. Therefore for the practical purposes it is important to investigate how the chaotic wave-current field dynamics can be modeled by the extended NLSE of Smith. In order to investigate this behavior we generate a chaotic wave-current field using a similar but different technique to the ones discussed by Akhmediev et.al. (2009a) and Bay\i nd\i r (2015). 

\noindent {\bf{II. 1. A split-step scheme for the NLSE of Smith}}

One of the very widely used class of numerical techniques in computational mathematics is the spectral methods. Some applications of the spectral methods is presented by Bay\i nd\i r (2009, 2015b, 2015c, 2015d) and Karjadi et. al. (2010, 2012). Their broader discussion is presented by Trefethen (2000). In spectral methods the spatial derivatives are handled by the orthogonal transformation techniques. The Fourier transform is the most popular choice for the periodic domains (Bay\i nd\i r; 2015, 2015e, 2015f). The temporal derivatives in the governing equations is evaluated by time integrating schemes such as Adams-Bashforth and Runge-Kutta etc. (Trefethen, 2000; Demiray and Bay\i nd\i r, 2015). 

One of the very popular Fourier spectral schemes with efficient time integration is the split-step Fourier method (SSFM). In the SSFM, the time integration is performed by time stepping of the exponential function for the governing equation with a first order time derivative (Bay\i nd\i r, 2015). For the extended NLSE of Smith we implement a split-step Fourier scheme. We take first part of the extended NLSE of Smith for $\alpha=0.5, \beta=1, \gamma=1$ as
\begin{equation}
i\psi_t= -\left| \psi \right|^2\psi + x \left| \frac{dU}{dx} \right| \psi
\label{eq13}
\end{equation}
which can be solved as
\begin{equation}
\tilde{\psi}(x,t_0+\Delta t)=e^{i \left( \left| \psi(x,t_0)\right|^2-x \left|dU/dx \right| \right) \Delta t}\ \psi(x,t_0)
\label{eq14}
\end{equation}
where $\Delta t$ denoted the time step. The remaining part of the extended NLSE of Smith is written as
\begin{equation}
i\psi_t=-\frac{1}{2}\psi_{xx}
\label{eq15}
\end{equation}
Using the Fourier series we obtain
 \begin{equation}
\psi(x,t_0+\Delta t)=F^{-1} \left[e^{-ik^2\Delta t/2}F[\tilde{\psi}(x,t_0+\Delta t) ] \right]
\label{eq16}
\end{equation}
where $k$ is the wavenumber parameter. Combining Eq.\,(\ref{eq14}) and Eq.\,(\ref{eq16}), the complete form of the SSFM scheme for the extended NLSE of Smith can be written as
 \begin{equation}
\psi(x,t_0+\Delta t)=F^{-1} \left[e^{-ik^2\Delta t/2}F[ e^{i\left( \left| \psi_0\right|^2-x \left|dU/dx \right| \right) \Delta t}\ \psi_0 ] \right]
\label{eq17}
\end{equation}
where $\psi(x,t_0)=\psi_0$. Starting from the chaotic initial conditions described below Eq.\,(\ref{eq18}) and Eq.\,(\ref{eq19}), we obtain the numerical solution of the extended NLSE of Smith for later times by the SSFM. This form of the SSFM requires two fast Fourier transform operations per time step. The number of spectral components are taken as $N=2048$ for using the fast Fourier transforms efficiently. The time step is selected as $\Delta t=0.05$ which does not cause any stability problems.

\noindent {\bf{II.2. Initialization of the chaotic wave field}}

We start simulations using a wave with constant amplitude of 1 and using a chaotic perturbation in the ocean current profile. Such a sea state becomes unstable and it evolves into a full-scale chaotic wave field.  For this purpose we introduce the initial conditions
\begin{equation}
\left|\psi \right|_0=1
\label{eq18}
\end{equation}
and
\begin{equation}
\frac{dU}{dx} (x,t)=\mu+\alpha [r_1+i r_2]
\label{eq19}
\end{equation}
where $i$ is the imaginary number, $\mu$ and $\alpha$ are some constants which determine the magnitude of chaotic perturbations and $r_1$ and $r_2$ are uniformly distributed random number vectors with values in the interval of [-1,1]. It is known that chaotic perturbations in the current profile imposes chaotic perturbations in the ocean surface wave field and vice versa. Therefore we use the expression in (\ref{eq19}) not only as an initial condition but also impose it in time to model this phenomena. The chaotic wave field with these initial conditions evolves into a wave field which exhibits many amplitude peaks, with some of them becoming rogue waves, similar to the results presented by Akhmediev et.al. (2009a, 2011) and Bay\i nd\i r (2015). It is possible to add perturbations to the current profile with a characteristic current length scale $L_{pert}$, by multiplying the second term in the (\ref{eq19}) by a factor of $\exp(i 2 \pi x / L_{pert})$. Or one can add perturbations with different current length scale using Fourier analysis. However in the present study for illustrative purposes we do not consider such as scale. The actual water surface fluctuation for this initial condition would be given by $ \left|\psi\right| \exp{[i\omega t]}$ where $\omega$ is some carrier wave frequency however we only consider $\left|\psi\right|$ in our numerical simulations.  

\section{Results and Discussion}

\begin{figure}[h]
\begin{center}
   \includegraphics[width=3.4in]{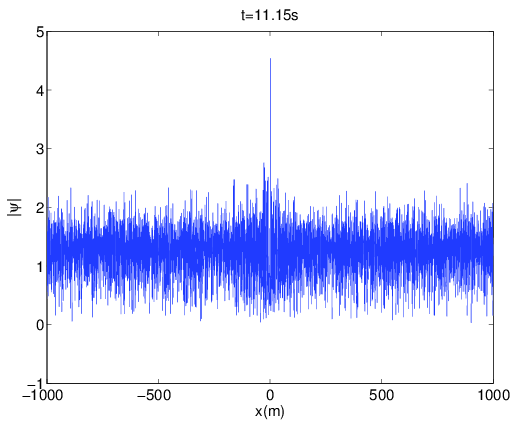}
  \end{center}
\caption{\small Rogue wave in the chaotic wave field.}
  \label{fig1}
\end{figure}

In Fig.\,\ref{fig1}., a typical run of the numerical simulations for the chaotic wave field is presented. A rogue wave is appearing around $x \approx 2m$. To model a uniform current with a chaotic perturbation, the parameters $\mu$ and $\alpha$ of Eq.\,(\ref{eq19}) are selected as $0$ and $0.2$, respectively. The uniform wave field with this perturbation rapidly turns into a chaotic sea state (Fig.\,\ref{fig1}). The actual spatial domain of the numerical simulations is as long as $L=[-500, 500]$. The peak amplitude for the appearing rogue wave in this simulation is 4.54 hence it can not be described by the first order rational soliton  solution (Akhmediev et.al., 2009a). Therefore we compare it with the second order rational soliton solution of NLSE. 

\

\begin{figure}[h]
\begin{center}
   \includegraphics[width=3.4in]{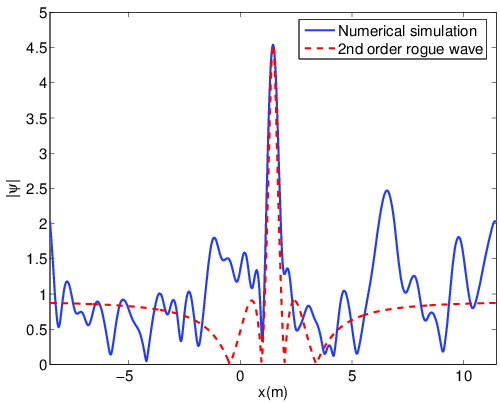}
  \end{center}
\caption{\small Comparison of the numerical simulation ({\color{blue} ---}) with the analytical second order rational soliton ({\color{red} -.-}) for the rogue wave profile.}
  \label{fig2}
\end{figure}

In order to analyze the shape of the rogue wave, in the Fig.\,\ref{fig2} we present a comparison of the rogue wave obtained in numerical simulations with the second order rational soliton solution defined by Eq.\,(\ref{eq05}). In this plot, the continuous blue line represents the numerical simulations. The exact second order rational soliton solution is shown by the dashed red line. The peak value for the analytical second order rational rogue wave is 5.00. Therefore we use a scaling factor of $B=4.54/5.00=0.908$ in the scaling law defined by Eq.\,(\ref{eq09}) and perform comparisons accordingly. The central part of the rogue wave accurately follows the exact profile. The discrepancy in the tails of the peak are due to random smaller amplitude waves of the chaotic field that are in the vicinity of the peak. It is possible to conclude that the profile around the peak closely follows the second-order rational solution. The rogue waves in the form the second order rational soliton can also be described by the collision of Akhmediev breathers (Akhmediev et.al., 2009a). Thus the rogue wave solutions of the extended NLSE of Smith can be described by collusions of such breathers as well.

\begin{figure}[h]
\begin{center}
   \includegraphics[width=3.4in]{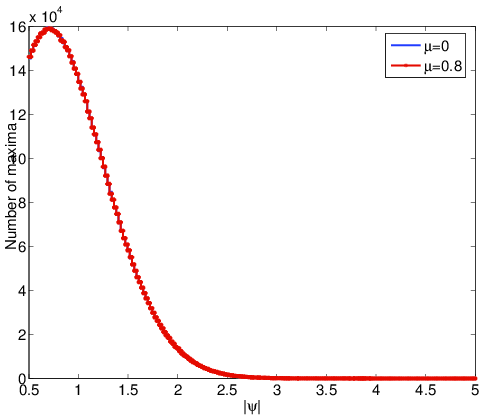}
  \end{center}
\caption{\small Histogram of the maxima appearing in simulations of the chaotic wave field for $\alpha=0.2$,  $\mu=0$  ({\color{blue} ---})  vs. $\mu=0.8$  ({\color{red} -.-})}
  \label{fig3}
\end{figure}

\begin{figure}[h]
\begin{center}
   \includegraphics[width=3.4in]{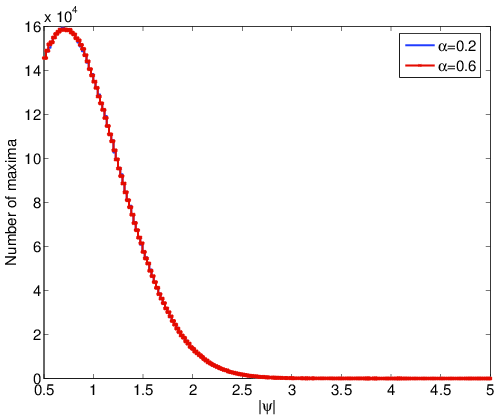}
  \end{center}
\caption{\small Histogram of the maxima appearing in simulations of the chaotic wave field for $\mu=0$,  $\alpha=0.2$  ({\color{blue} ---})  vs. $\alpha=0.6$  ({\color{red} -.-})}
  \label{fig4}
\end{figure}

In order to quantify the probabilities of occurrence for different values of the maxima, two different initial conditions are considered. In the first case we take $\alpha=0.2$ and $\mu=0.0$ or $\mu=0.8$. In the second case we take $\mu=0$ and $\alpha=0.2$ or $\alpha=0.6$. The histograms for this first case are depicted in Fig.\,\ref{fig3} and the histograms for the second case are depicted in Fig.\,\ref{fig4}. For these two cases, we plotted the number of maxima appearing in each simulation within small fixed intervals of amplitude. This intervals are selected as $0.05$. We only count maxima above a lower limit of 0.5. The total number of maxima in each case is equal to 6.4 million. This massive runs are performed in wide temporal intervals as well as with long propagation distance $x$ to ensure statistical convergence. Each simulation was repeated 50 times with new initial conditions with same values of $\mu$ and $\alpha$. These massive simulations allowed us to obtain several thousands of maxima within each small amplitude segment. We observe only a few maxima bigger than 4.5.

These results clearly show that the maxima of the chaotic wavefield is independent from $\mu$ and  $\alpha$, the mean slope of the current gradient and the magnitude of the chaotic perturbations given in Eq.\,(\ref{eq19}). This was expected because the last term in the extended NLSE of Smith given in Eq.\,(\ref{eq01}) accounts for the phase shifts due to wave-current interaction. It does not amplify the chaotic wave field of the cubic NLSE. The average energy for both of the cases are $E \left[ \left|\psi \right|^2 \right]=1.25$, to the second digit accuracy in decimal points. This result, together with the histograms confirm that the last term in the extended NLSE of Smith does not change the total energy of the chaotic wave field. It rather behaves like a phase lag term, altering the location of occurrence of rogue waves. This can also be verified analytically; the transformation given in Eq.\,(\ref{eq02}) only introduces a shift to the solutions of the cubic NLSE.

\section{Conclusion}
In this study we have developed a numerical framework to study how the chaotic perturbations in the ocean current profile lead to the development of the rogue waves on the ocean surface. For this purpose we have used an extended nonlinear Schr\"{o}dinger equation (NLSE) developed by Smith (1976). This extended NLSE takes the current gradient effects into account. We have implemented a split-step scheme and we have numerically showed that the extended NLSE of Smith is unstable against random chaotic perturbations in the current profile. Therefore the initial monochromatic wave field turns into a chaotic sea state with many apparent peaks. We have numerically showed that rogue waves appearing on the ocean surface due to perturbations in the ocean current profile are in the form of rational soliton solutions of the cubic NLSE. We have also discussed the effects of magnitude of the chaotic current profile perturbations on the statistics of the rogue wave generation on the ocean surface. We have showed that extension term in the NLSE of Smith shifts the phases of the solutions of the NLSE, thus locations of the rogue wave peaks.




\end{document}